MECHANISM OF AMBIPOLAR FIELD-EFFECT TRANSISTORS ON ONE-DIMENSIONAL ORGANIC MOTT INSULATORS

K. Yonemitsu[1,2]

[1]Institute for Molecular Science, Okazaki 444-8585, Japan. E-mail: kxy@ims.ac.jp
[2]Graduate University for Advanced Studies, Okazaki 444-8585, Japan

1  INTRODUCTION

Field-effect transistors are fabricated on many molecular materials. Their characteristics are not always explained in the conventional manner. Indeed, ambipolar characteristics are reported in metal-insulator-semiconductor field-effect transistor device structures based on organic single crystals of the quasi-one-dimensional Mott insulator (BEDT-TTF)($F_2$TCNQ) [BEDT-TTF=bis(ethylenedithio)tetrathiafulvalene, $F_2$TCNQ=2,5-difluorotetracyanoquinodimethane].[1] In quasi-one-dimensional systems with coherent band transport, the insulator-electrode interface barrier potentials are crucial to the current-voltage characteristics. For example, carbon nanotube field-effect transistors operate as Schottky barrier transistors, in which transistor action occurs primarily by varying the contact resistance rather than the channel conductance.[2,3] The Schottky barriers are sensitive to the work-function difference between the channel and the source/drain electrodes. As long as the work functions of the electrodes are different from that of the channel, the characteristics are unipolar in general. By matching their work functions, the ambipolar field-effect characteristics are achieved.[2,3]

Thus, the ambipolar characteristics of the organic Mott insulator[1] imply that electron correlations are crucial. In organic insulators and conductors, the importance of electron correlations is acknowledged in the equilibrium phase diagram and the bulk transport properties.[4] For example, the motion of carriers is collective and sometimes confined into chains, which is qualitatively different from the conventional behavior of individual quasiparticles. Because of the collective nature, the motion of carriers inside the Mott insulator is correlated with that near the insulator-electrode interface. In order to clarify the relationship between the electron correlations and the Schottky barriers, we calculate the current-voltage characteristics using the one-dimensional Hubbard model for Mott insulators and adding to it potentials that originate from the long-range Coulomb interaction and are modified by the work-function difference, the gate bias and the drain voltage. The time-dependent Schrödinger equation is combined with the Poisson equation and numerically solved self-consistently at each site and time within the unrestricted Hartree-Fock approximation. The ambipolar field-effect characteristics are shown to be caused by balancing the correlation effect with the barrier effect. For the gate-bias polarity with higher Schottky barriers, the correlation effect is weakened accordingly.

## 2 METHOD AND RESULTS

### 2.1 One-Dimensional Model for Field-Effect Transistors

Many works employ the one-dimensional Hubbard model for a Mott insulator, to which the tight-binding model is attached for metallic electrodes.[5,6] We add a scalar potential $v_l$,[7]

$$H = \sum_l (\varepsilon_l + v_l) n_l - \sum_{l,\sigma} \left( t_{l,l+1} c^+_{l,\sigma} c_{l+1,\sigma} + \text{h.c.} \right) + \sum_l U_l (n_{l\uparrow} - 1/2)(n_{l\downarrow} - 1/2), \tag{1}$$

where $c^+_{l,\sigma}$ ($c_{l,\sigma}$) creates (annihilates) an electron with spin $\sigma$ at site $l$, $n_{l,\sigma} = c^+_{l,\sigma} c_{l,\sigma}$, and $n_l = \sum_\sigma n_{l,\sigma}$. The site energy $\varepsilon_l$ is set at 0 in the crystal and at $\phi$ in the electrodes. The absolute value of the transfer integral $|t_{l,l+1}|$ is set at $t_c$ if either $l$ or $l+1$ is in the crystal and at $t_e$ otherwise. The on-site repulsion $U_l$ is set at $U$ in the crystal and at 0 in the electrodes. When we consider band insulators instead, we use $U = 0$ and replace $t_c$ by $t_c - (-1)^l \delta t$. The total number of electrons is the same as the number of sites. The left and right electrodes are regarded as the source and drain, respectively. The periodic boundary condition is imposed by introducing the Peierls phase, which is proportional to the vector potential, to the transfer integral for finite drain voltage $V_D$. The scalar potential $v_l$ obeys the Poisson equation on the discrete lattice ($l=1, 2, \ldots L$),

$$v_{l+1} - 2v_l + v_{l-1} = -V_{Pl} (\langle n_l \rangle - n_{Bl}), \tag{2}$$

where the parameter $V_{Pl}$ comes from the long-range Coulomb interaction and $n_{Bl}$ from the background charge. The parameter $V_{Pl}$ is set at $V_{Pc}$ in the crystal and at $V_{Pe}$ in the electrodes, with $V_{Pe} \ll V_{Pc}$. The gate bias $U_G$, which is relative to the middle of the source-drain potential,[1] is assumed to appear with the gate efficiency factor $\alpha$ in the boundary value of the Poisson equation,

$$v_0 - v_{L/2} = v_L - v_{L/2} = \alpha U_G - \phi, \tag{3}$$

where $l=L/2$ is set at the middle of the crystal and $l=0$ ($L$) in the source (drain) electrode is the furthest point from the interfaces. Hereafter we rewrite $\alpha U_G$ as $U_G$ for simplicity.

The time-dependent Schrödinger equation is numerically solved in the unrestricted Hartree-Fock approximation simultaneously with the Poisson equation. The self-consistency is imposed at each time and site in these equations. The drain current $I_D$ is obtained by averaging the current density,[8]

$$J(t) = -(1/L) \sum_{l,\sigma} \left[ it_{l+1,l}(t) c^+_{l+1,\sigma} c_{l,\sigma} - it_{l,l+1}(t) c^+_{l,\sigma} c_{l+1,\sigma} \right], \tag{4}$$

over the period $0 < t < \Delta t$ with $\Delta t = 2\pi L/(4V_D)$. This definition leads to positive (negative) $I_D$ for positive (negative) $V_D$ in general. For calculations, we set $t_e = t_c = 1$, $V_{Pc} = 0.05$, $V_{Pe} = 10^{-3}$, and $L=100$ with 51 crystal sites. Because of finite $V_{Pe}$, a charge is induced at the metallic surface. Its detail is beyond the scope of the Poisson equation, but it does not affect the conclusion. Both $U=2$ used for the Mott insulator and $\delta t=0.17$ used for the band insulator give a charge gap of about $\Delta_{CG}=0.68$. The work-function difference $\phi$ is set at $-1.5$, so that the electrodes have a higher work function than the crystal. Then, the Schottky barriers appearing at the insulator-electrode interfaces become higher for the electron injections $U_G>0$ and lower for the hole injections $U_G<0$.

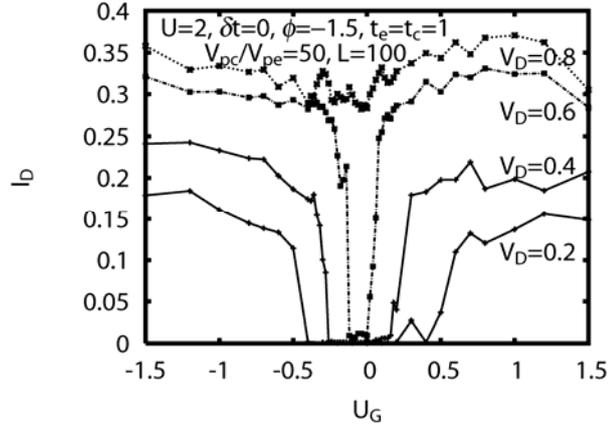

**Figure 1** $I_D$-$U_G$ characteristics at various $V_D$ for the Mott insulator with $U=2$, $\delta t=0$ and the work-function difference $\phi=-1.5$. The other parameters are $t_e=t_c=1$, $V_{Pc}/V_{Pe}=50$, and $L=100$.

The $I_D$-$U_G$ characteristics of the Mott insulator are nearly symmetric (i.e., even functions) with respect to the polarity of $U_G$ (Figure 1). This is the case even if the work-function difference makes the Schottky barriers higher for $U_G>0$. For small $V_D$ and $|U_G|$, $I_D$ is suppressed by the charge gap, and $I_D$ increases with $U_G>0$ and with $U_G<0$ in a very similar manner. These ambipolar characteristics are achieved, for the gate-bias polarity with the higher Schottky barriers ($U_G>0$), by weakening the correlation: the deviation from half filling inside the insulator is larger, reducing the umklapp scattering strength.

The $I_D$-$U_G$ characteristics of the band insulator are very asymmetric with respect to the polarity of $U_G$ (Figure 2). This is reasonable, from the viewpoint of individual particles, because the high Schottky barriers for $U_G$>0 simply enhance the backward scatterings at the insulator-electrode interfaces and consequently reduce the current density for $U_G$>0. For further electron injections, $I_D$ remains suppressed, so that the characteristics are unipolar. Thus, the field-effect characteristics depend largely on the origin of the charge gap in the insulator. The potential distribution and the local density of states in the band insulator are similar to those in the Mott insulator as long as the gate bias is so small that the current does not flow. Therefore, the differences are manifested only in the dynamical and non-equilibrium condition.

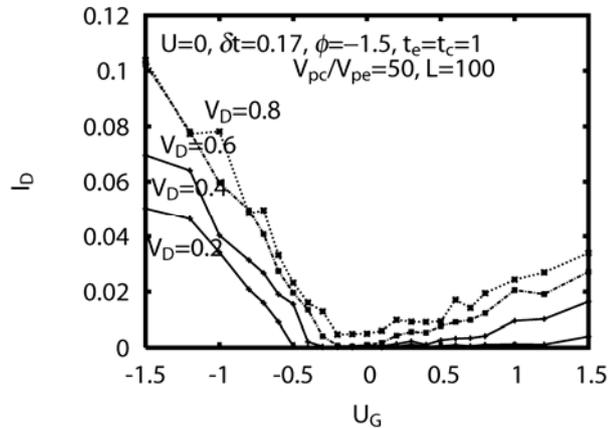

**Figure 2** $I_D$-$U_G$ characteristics at various $V_D$ for the band insulator with $U=0$, $\delta t=0.17$ and the work-function difference $\phi=-1.5$. The other parameters are the same as in Figure 1.

## 2.2 Insulators Attached to Electrodes with Different Work Functions

In the model above, the gate electrode modulates only the potential at the midpoint of the crystal. Therefore, it is quite reasonable for one to regard only one insulator-electrode interface as essential to the peculiar characteristics. Namely, the above field-effect characteristics would be approximately given by the difference between two current-voltage relations for systems with one interface each. However, two interfaces are needed for the current to be measured. Here, we consider insulators attached to two electrodes with different work functions, where only one insulator-electrode interface gives a significantly high Schottky barrier.

We employ the model (1), where the site energy $\varepsilon_l$ is set at 0 in the crystal, at $\phi_L$ in the left electrode (for small $l$'s), and at $\phi_R$ in the right electrode (for large $l$'s). Otherwise common parameters are assigned to the two electrodes. The periodic boundary condition is imposed again by introducing the Peierls phase to the transfer integral for finite voltage $V$, which corresponds to $V_D$ in the field-effect transistors. Now, the boundary values of the Poisson equation are set by

$$v_0 - v_{L/2} = -\phi_L, \qquad (5)$$

which compensates for the work-function difference between the insulator and the left electrode, and by

$$v_L - v_{L/2} = -\phi_R, \qquad (6)$$

which compensates for the work-function difference between the insulator and the right electrode. The definition of the current $I$ is the same as that of $I_D$ in the field-effect transistors.

We numerically calculate the time-evolution of the system with the same parameters as before unless otherwise noted. Both $U$=2 used for the Mott insulator and $\delta t$=0.17 used for the band insulator give a charge gap of about $\Delta_{CG}$=0.68; $U$=1.9 and $\delta t$=0.145 give $\Delta_{CG}$=0.58; and $U$=1.8 and $\delta t$=0.1225 give $\Delta_{CG}$=0.49. The parameter $\phi_R$ is so set that the Fermi level of the right electrode, if isolated, is set at the top of the lower Hubbard band or the valence band. The parameter $\phi_L$ is so set that the Fermi level of the left electrode, if isolated, is set at about 3/4 times the charge gap higher than the bottom of the upper Hubbard band or the conduction band. Then, only the Schottky barrier at the left interface is significantly high. It becomes higher for the left-going current $V$<0 and lower for the right-going current $V$>0.

The $I$-$V$ characteristics of the Mott insulator are rather anti-symmetric (i.e., odd functions) with respect to $V$ (Figure 3). This is the case even if the work-function difference at the left interface makes the Schottky barrier higher for $V<0$. For small $V$, $|I|$ is suppressed by the charge gap, and $|I|$ increases with $V<0$ and with $V>0$ in a similar manner above the threshold. For $V<0$, the correlation is accordingly weakened.

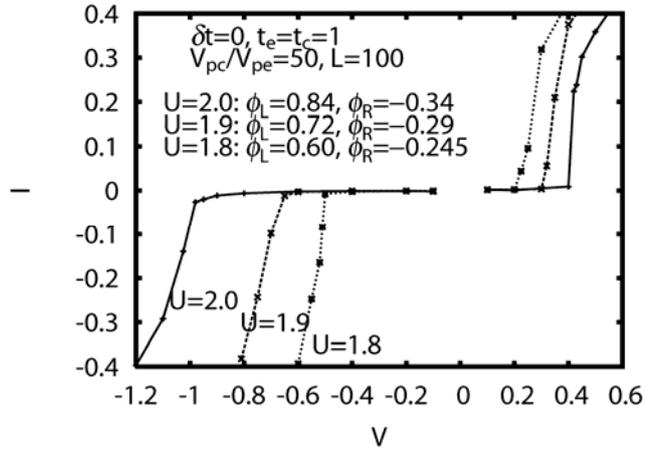

**Figure 3** *I-V characteristics for Mott insulators with various U and $\delta t=0$. The work-function differences $\phi_R$ and $\phi_L$ are such that the Fermi level of the right electrode, if isolated, is set at the lower Hubbard band and that of the left electrode at about 3/4 times the charge gap higher than the upper Hubbard band. The other parameters are the same as in Figure 1.*

The *I-V* characteristics of the band insulator are far from anti-symmetric with respect to $V$ (Figure 4). Simply because the Schottky barrier at the left interface is higher for $V<0$, the enhanced backward scattering suppresses the current density for $V<0$ only. In short, the field-effect characteristics can be interpreted as the difference between the current-voltage relation governed by the insulator-source-electrode interface and that by the insulator-drain-electrode interface ($V_D$ times the derivative in the limit of small $V_D$).

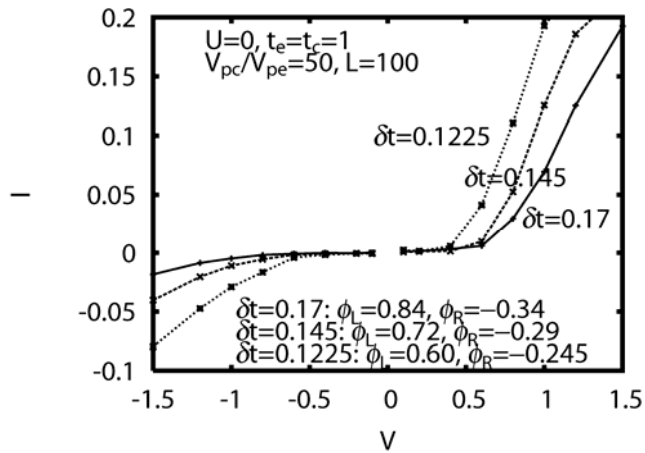

**Figure 4** *I-V characteristics for band insulators with various $\delta t$ and $U=0$. The work-function differences $\phi_R$ and $\phi_L$ are such that the Fermi level of the right electrode, if isolated, is set at the valence band and that of the left electrode at about 3/4 times the charge gap higher than the conduction band. The other parameters are the same as in Figure 1.*

## 3 CONCLUSION

In the one-dimensional Hubbard model for a Mott insulator attached to the tight-binding model for electrodes, both with scalar and vector potentials for Schottky barriers at interfaces, the field-effect characteristics are calculated by solving the time-dependent Schrödinger and Poisson equations simultaneously and self-consistently. They are always ambipolar for Mott insulators even if the work function of the crystal is quite different from that of the electrodes. This is in contrast to the generally unipolar characteristics for band insulators. In Mott insulators, for gate-bias polarity with the higher Schottky barriers, compared with the opposite polarity, we find that the deviation from half filling inside the insulator is larger, which reduces the umklapp scattering strength. What does this mean? Electrons are scattered backwards at the insulator-electrode interface. If they behaved individually as in band insulators, the charge-density distribution would be largely altered at high cost owing to the on-site repulsion. To avoid this situation, all the electrons are so correlated that they flow collectively. This is made possible by the fact that the strong backward scattering at the interface is counterbalanced by weakening the umklapp scattering inside the insulator. Therefore, the robust ambipolar field-effect characteristics in Mott insulators are due to balancing the correlation effect with the barrier effect, which allows the collective charge transport. The ambipolar (i.e., nearly even functions with respect to $U_G$) field-effect characteristics can be explained by the difference between the anti-symmetric (i.e., nearly odd functions with respect to $V$) $I$-$V$ characteristics at the two insulator-electrode interfaces. The latter is a sufficient condition for the former, and indeed theoretically realized for Mott insulators attached to electrodes with quite different work functions, where only one insulator-electrode interface gives a significantly high Schottky barrier. The qualitative differences between the non-equilibrium properties of Mott insulators and those of band insulators are also found in photoinduced phase transitions between these insulators: the dynamics are cooperative in Mott insulators[9] and uncooperative in band insulators.[10]

**Acknowledgements**

The author is grateful to T. Hasegawa for showing his data prior to publication and for enlightening discussions including systems attached to electrodes with different work functions. This work was supported by Grants-in-Aid and NAREGI Nanoscience Project from the Ministry of Education, Culture, Sports, Science and Technology, Japan.